\documentclass[reprint, superscriptaddress,
amsmath,amssymb,
aps,
prl,
]{revtex4-1}
\pdfoutput=1
\usepackage{graphicx}
\usepackage{dcolumn}
\usepackage{bm}
\usepackage{siunitx}
\DeclareSIUnit\gauss{G}
\usepackage{braket}
\usepackage[normalem]{ulem}
\usepackage[dvipsnames]{xcolor}
\usepackage{hyperref}
\usepackage{tabularx}
\usepackage{braket}
\newcolumntype{Y}{>{\centering\arraybackslash}X}

\begin{document}


\title{Polaronic hybridization of atoms, dimers and trimers in a Bose-Einstein condensate}

\author{Carsten Robens}
\thanks{These authors contributed equally.}
\affiliation{
 MIT-Harvard Center for Ultracold Atoms, Research Laboratory of Electronics, and Department of Physics,
 Massachusetts Institute of Technology, Cambridge, Massachusetts 02139, USA\\
}%
\author{Arthur Christianen}
\thanks{These authors contributed equally.}

\affiliation{Institute for Theoretical Physics, ETH Zürich, 8093 Zürich, Switzerland}%
\affiliation{%
Max-Planck-Institut für Quantenoptik, 85748 Garching, Germany
}%
\affiliation{%
Munich Center for Quantum Science and Technology (MCQST), 80799 Munich, Germany
}%
\author{Alexander Y. Chuang}
\thanks{These authors contributed equally.}
\affiliation{
 MIT-Harvard Center for Ultracold Atoms, Research Laboratory of Electronics, and Department of Physics,
 Massachusetts Institute of Technology, Cambridge, Massachusetts 02139, USA\\
}%
\author{Huan Q. Bui}
\affiliation{
 MIT-Harvard Center for Ultracold Atoms, Research Laboratory of Electronics, and Department of Physics,
 Massachusetts Institute of Technology, Cambridge, Massachusetts 02139, USA\\
}%
\author{Yiming Zhang}
\affiliation{
 MIT-Harvard Center for Ultracold Atoms, Research Laboratory of Electronics, and Department of Physics,
 Massachusetts Institute of Technology, Cambridge, Massachusetts 02139, USA\\
}%

\author{Richard Schmidt}
\affiliation{
Institute for Theoretical Physics, Heidelberg University, Philosophenweg 16, 69120 Heidelberg, Germany \\
}%

\author{Martin Zwierlein}
\affiliation{
 MIT-Harvard Center for Ultracold Atoms, Research Laboratory of Electronics, and Department of Physics,
 Massachusetts Institute of Technology, Cambridge, Massachusetts 02139, USA\\
}%

\date{\today}
\begin{abstract}

The Bose polaron problem of an impurity immersed in a Bose-Einstein condensate (BEC) has been predicted to feature strong correlations arising from bound states of multiple bosons with the impurity. While direct experimental evidence has so far remained elusive, here we observe clear signatures of three-body correlations in Bose polarons. We perform radiofrequency spectroscopy on $^{40}$K impurities in a BEC of $^{23}$Na and identify polaronic hybrid states that can be understood as superpositions of the bare atom, a NaK dimer and a Na$_2$K trimer, coupled through coherent particle exchange with the condensate. We show that the main spectroscopic features are captured by a simple three-level model without free parameters. Our work shows how a condensate environment can coherently hybridize bound states of different composition and mass, reminiscent of quark-flavor mixing described by the Cabibbo-Kobayashi-Maskawa (CKM) matrix in particle physics.
\end{abstract}

\pacs{Valid PACS appear here}
\maketitle

A particle immersed in a quantum medium can become dressed by the bath's excitations into a polaron, a quasiparticle with modified energy and mass~\cite{landau:1933,pekar:1946,froelich:1954,alexandrov:2010}. Ultracold quantum gases provide an ideal platform to study polaron physics in the strong-coupling regime~\cite{grusdt:2025,massignan:2025}, because interactions are tunable~\cite{chin:2010} and impurities can form bound states with particles of the bath. 
For a fermionic bath, Pauli blocking limits the number of bath particles in close proximity to the impurity. The emergent Fermi polaron~\cite{Schirotzek2009, koschorreck:2012, scazza:2022,massignan:2025} is therefore well described as a superposition of the bare impurity and a single particle-hole excitation on top of the Fermi sea~\cite{chevy:2006, combescot:2007, Prokofev2008a}.
If the bath instead is bosonic, there is in principle no limit on the number of bosons that can bind to the impurity. At strong interactions, the resulting Bose polaron~\cite{rath:2013,ardila:2015,shchadilova:2016, grusdt:2025,massignan:2025} can thus feature higher-order correlations that make the problem rich but intricate~\cite{levinsen:2015,sun:2017a,sun:2017b,yoshida:2018,christianen:2022b,christianen:2024}.

Despite extended experimental efforts~\cite{hu:2016,jorgensen:2016,yan:2020,skou:2022,morgen:2025,etrych:2025,etrych:2025b,morgen:2026}, so far no clear evidence of higher-order correlations has been observed. It is challenging to spectroscopically study such strongly correlated states, since they typically have little wave function overlap with the often non-interacting initial states. At the same time, these many-body bound states have short lifetimes due to three-body recombination, which also hinders their preparation in equilibrium.

\begin{figure}[bh!]
    \centering
\includegraphics[width=0.97\columnwidth]{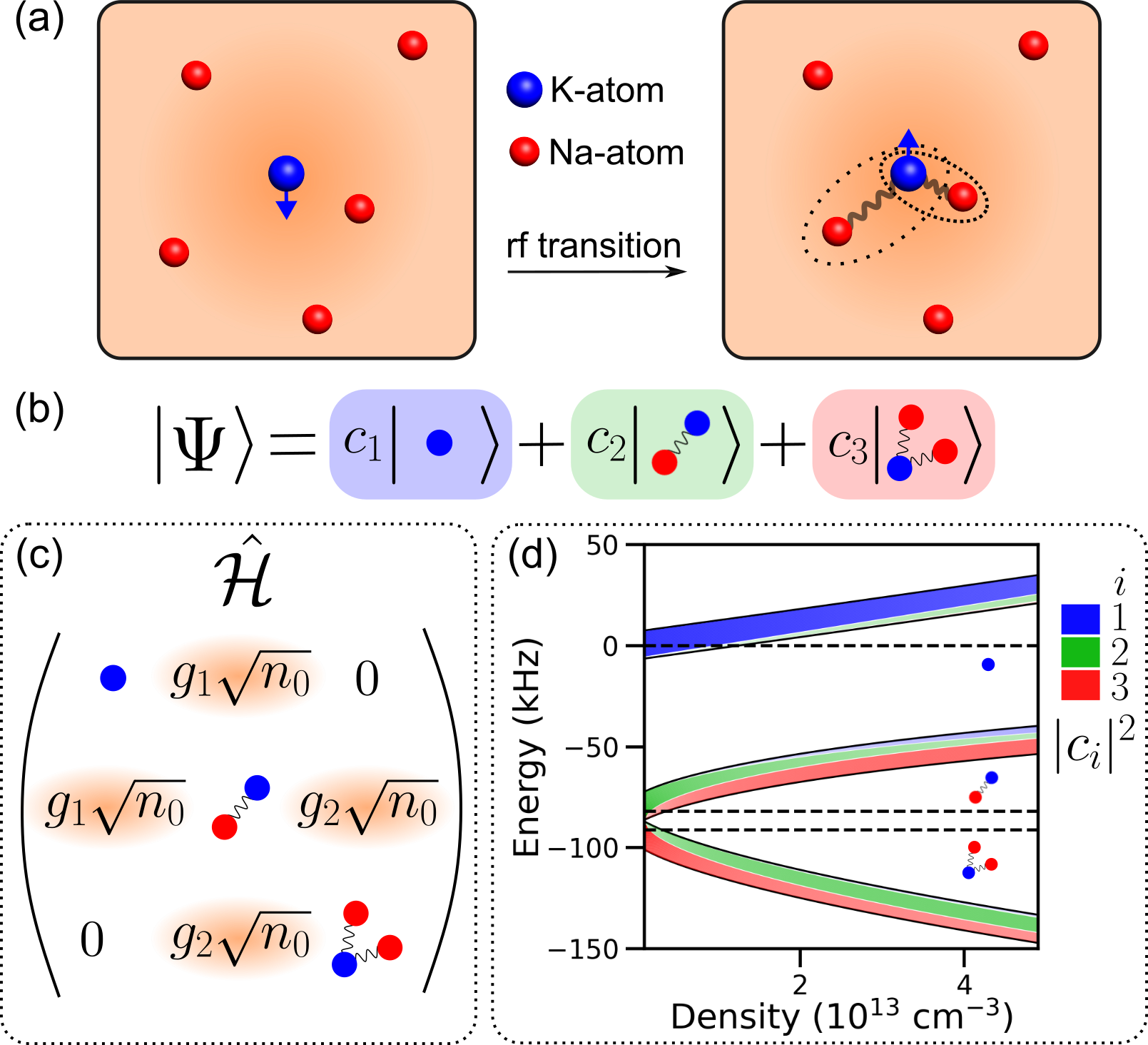}
    \caption{Emergence of atom-dimer-trimer superpositions. (a) The rf-association process from the initial-state K-atom immersed in a BEC of Na to the dimer-trimer hybridized state.  
    (b) Illustration of the wave function: a superposition of a free K-atom, a dimer, and a trimer. (c) The effective Hamiltonian in the basis of these three states. (d) The three eigenvalues of this Hamiltonian as a function of the BEC density $n_0$. The widths and opacities of the colored sub-lines represent the coefficients $|c_1|^2$ (blue), $|c_2|^2$ (green) and $|c_3|^2$ (red) for each of the eigenstates. The horizontal dashed lines indicate the energies of the three bare basis states.}
    \label{fig:intro}
\end{figure}

In this Letter, we report the observation of Bose polarons that can be described as superpositions of the bare atom with dimer and trimer states. We perform radiofrequency spectroscopy on $^{40}$K impurities in a Bose-Einstein condensate (BEC) of $^{23}$Na atoms (Fig.~\ref{fig:intro}(a)). In a thermal Na-K mixture, we recently identified a weakly bound Na$_2$K halo trimer state \cite{chuang:2025} lying close in energy to the NaK Feshbach dimer over an extended range of interaction strengths. We find that the role of this trimer state in the Bose polaron formed in the BEC can be understood through the simple three-level model shown in Fig.~\ref{fig:intro}(b)–(d). Because the impurity can coherently exchange particles with the BEC, the condensate induces an off-diagonal coupling between the dimer and trimer states. For our system the coupling by the BEC exceeds the small dimer-trimer splitting and produces strong level repulsion. The resulting two lowest eigenstates are nearly equal-weight superpositions of dimer and trimer states. The three levels of this effective model capture the main features in the observed spectra without free parameters, providing clear experimental evidence for polaronic hybridization of these few-body bound states.
The $3\times 3$ Hamiltonian describing the coupling of different particle sectors is reminiscent of the Cabibbo-Kobayashi-Maskawa (CKM) matrix~\cite{cabibbo:1963,kobayashi:1973,navas:2024}, which encodes quark-flavor mixing through interactions with W bosons.
Dimer and trimer superpositions have previously been created by an external radiofrequency drive in thermal gases~\cite{yudkin:2019}. Here, instead, it is the inherent coupling to the BEC that induces the hybridization, producing polaronic states that are fundamentally different in nature.

Our experiment starts with a Bose-Fermi mixture of $^{23}$Na and $^{40}$K at approximately 100\,nK, trapped in an optical dipole trap with Na trapping frequencies $2\pi \times (108, 112, 9)$ Hz. The BEC has a typical peak density of $n_\mathrm{B}\,{\approx}\,4 \times 10^{13}$cm$^{-3}$ while the Fermi gas has a two orders of magnitude lower density. 
As illustrated in Fig.~1(a), we perform radiofrequency-injection spectroscopy by driving $^{40}$K from its initial spin state, which we denote as $\ket{\downarrow}$, into a final spin state 
 $\ket{\uparrow}$ for which the impurity-boson interatomic potential supports a shallow bound state. By measuring the number of transferred atoms we can extract information about the presence of final states at the probed energy and, according to Fermi's Golden rule, about the overlap between these final states and the initial state.

An essential feature of our experiment is that we boost the overlap with the molecular final states of interest by starting with an interacting initial state $\ket{\downarrow}$. This is possible because the $\ket{\downarrow}$ and $\ket{\uparrow}$ states have overlapping Feshbach resonances, see Fig.~\ref{fig:Feshbachplots} in the Supplementary Material (SM). The initial-state scattering length is approximately constant in the probed magnetic field range and given by $(k_n a)^{-1} \sim -1$, where $k_n = (6\pi n_\mathrm{B})^{1/3}$ is the inverse interboson distance. Before performing spectroscopy, we thus allow the impurity to form an attractive Bose polaron by letting it thermalize with the bath for 25ms, increasing the density of the BEC close to the impurity. This enables us to drive the rf-transition to final bound states faster than the decay of the short-lived final states.  %

\begin{figure}[ht!]
    \centering
    \includegraphics[width=0.95\columnwidth]{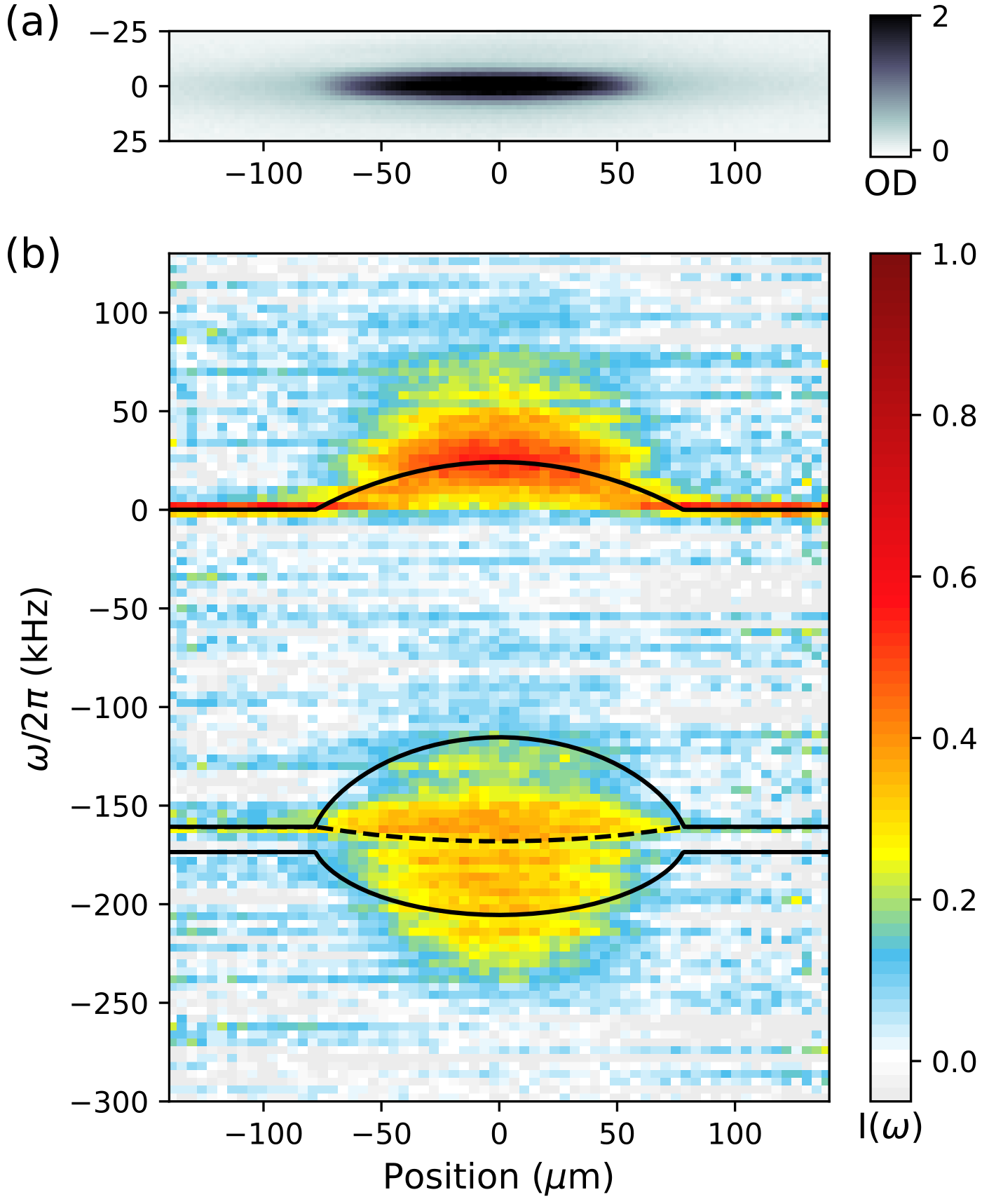}
    \caption{Radiofrequency spectra of $^{40}$K impurities strongly depend on their position in the $^{23}$Na BEC. (a) Density profile of the BEC, where the grayscale indicates the optical density (OD). (b) Spatially resolved spectra at $B=102.5$G of initial state $\ket{\downarrow}$ depletion when driving at given detuning from the $\ket{\downarrow}$-$\ket{\uparrow}$ bare hyperfine transition. The solid lines indicate the theory predictions from the three-level model, whereas the dashed line shows the theory result without three-body correlations.}
    \label{fig:halfMoon}
\end{figure}

We observe the effect of the background BEC on the impurity atoms by spatially resolving the rf-signal at different positions in the atomic cloud, as shown in Fig.~\ref{fig:halfMoon} for the magnetic field of $B=102.5\,\rm G$. For each position, we obtain the signal integrated across the narrow dimension of the cigar-shaped atom cloud (see Fig.~\ref{fig:halfMoon}(a)). 
The  black lines correspond to the fit-parameter-free theoretical predictions using the three-level model as depicted in Fig.~\ref{fig:intro}(b), which will be discussed below.

 At low density, i.e., near the edge of our cloud, we observe the bare free-to-free and free-to-bound transitions, see Fig.~\ref{fig:expcuts} (SM). Here we reproduce the dimer and trimer features we have observed previously in thermal clouds \cite{chuang:2025}.  
 
 When the density is increased towards the center of the cloud, the narrow peaks observed at low density shift and broaden dramatically due to the coupling with the BEC. The upper peak shifts upward in energy, due to the formation of a repulsive polaron~\cite{scazza:2022}. The lower peak shifts on average to lower energies, and its line shape changes strongly, evolving into a broad feature. The coupling with the BEC medium is so strong that the lowest energy at which we observe significant signal lies approximately 50 kHz below both the dimer and the trimer energies; that is about a third of the dimer binding energy. We are therefore clearly in the strong-coupling regime, where the polaronic correlations are equally important as molecular binding.

We show theoretically that the large energy shift and broadening as a function of the density can be explained as originating from the atom-dimer-trimer hybridization. To this end, we consider a variational wave function as shown in Fig.~\ref{fig:intro}(b) which consists of a superposition of a free-impurity, a dimer and a trimer state \cite{levinsen:2015,yoshida:2018,christianen:2024}. We use a microscopic model with model potentials which reproduce the low-energy scattering properties extracted from first-principles scattering calculations (see SM). The dimer and trimer wave functions are then variationally optimized to minimize the energy in the presence of the BEC. The effective three-level model is constructed by computing the matrix elements of the Hamiltonian in the basis of the free-impurity, dimer, and trimer states obtained from the microscopic calculation. The lowest-energy state of the three-level model is, by construction, the ground state of the variational calculation. 

The coupling between the different particle number sectors follows from displacing the bosonic creation and annihilation operators $\hat{b}^{\dagger}_{\bm{r}}$, $\hat{b}_{\bm{r}}$ (respectively creating and annihilating bosons at position $\bm{r}$), by the field $\phi=\sqrt{n_0}$ of the background BEC in the boson-impurity density-density interaction:
\begin{align} \label{eq:HBI}
    \hat{\mathcal{H}}_{\rm BI} &= \int d^3r d^3r' V(\bm{r}-\bm{r'}) \hat{b}^{\dagger}_{\bm{r}}\hat{b}_{\bm{r}} \hat{n}_{\rm I}(\bm{r'}), \\  \hat{U}^{\dagger}_{\phi}\hat{\mathcal{H}}_{\rm BI} \hat{U}_{\phi}&=\int d^3r d^3r' V(\bm{r}-\bm{r'}) (\phi^{\ast}+\hat{b}_{\bm{r}}^{\dagger}) (\phi+\hat{b}_{\bm{r}}) \hat{n}_{\rm I}(\bm{r'}).\nonumber
\end{align}
Here, $V(\bm{r}-\bm{r'})$ is the boson-impurity interaction potential and $\hat{n}_{\rm I}(\bm{r'})$ is the impurity density operator. The unitary $\hat{U}_{\phi}$ is the displacement operator. The quadratic $\hat{b}^{\dagger} \hat{b}$ term enables the formation of impurity-boson bound states, while the linear $\hat{b}^{\dagger}$ and $\hat{b}$ terms give rise to the off-diagonal terms in the Hamiltonian in Fig.~\ref{fig:intro}(c). The corresponding coefficients $g_1$ and $g_2$, as shown in the matrix in the figure, follow directly from the potential and the dimer and trimer wave functions (see SM). 

We find that for the experimental BEC densities, the ground state of this Hamiltonian contains the dimer and trimer contributions with comparable weights. This is illustrated in Fig.~\ref{fig:intro}(d), where the nature of the eigenstates is shown as a function of the BEC density. This striking effect arises because the coupling by the BEC on the off-diagonal of the three-level Hamiltonian is larger than the intrinsic dimer-trimer splitting. 

The splitting between the lowest two lines of the three-level model well describes the width of the experimental attractive polaron feature (see Fig.~\ref{fig:halfMoon}(b)). This indicates that the dimer-trimer hybridization sets the dominant energy scale determining the density-dependent shift and broadening of the attractive polaron branch. For reference, we also include a theoretical curve (dashed) where only a single excitation from the medium is admitted, as in the standard Chevy Ansatz ~\cite{chevy:2006}. This model, which by design cannot describe three-body correlations, predicts much weaker energy shifts than those observed experimentally. 

To further justify the interpretation of these states as dimer-trimer superpositions, we verify that the variationally optimized two- and three-body wave functions in the presence of the BEC still resemble the in-vacuum dimer and trimer wave functions. We find that for the parameters in Fig.~\ref{fig:halfMoon}(b) this overlap is around 90\%.

\begin{figure}[t]
	\includegraphics[width=0.95\columnwidth]{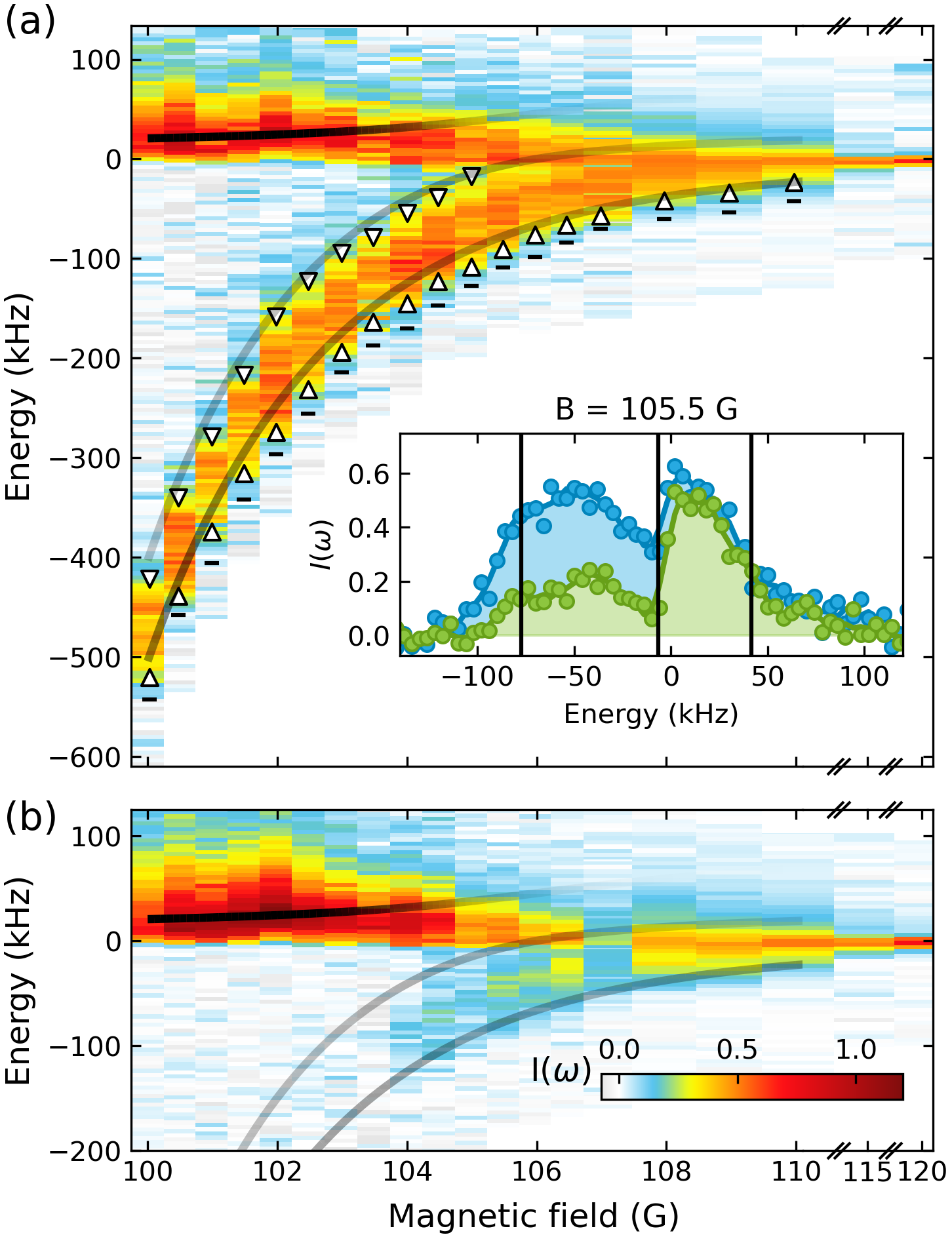}
	\caption{\label{fig:E_vs_B} Rf impurity spectra in the center of the BEC as a function of magnetic field near the $\ket{\uparrow}$ impurity-boson Feshbach resonance. (a) depletion of state $\ket{\downarrow}$ and (b) arrival of state $\ket{\uparrow}$.
    Rf pulse time is chosen at each field for $\sim 50\%$ depletion on the attractive branch, and spectra are normalized for equal transfer at the bare dimer energy. Inset: rf signal at $105.5\,\rm G$ (blue: depletion, green: arrival). In (a) the full width at half maximum of the signal is indicated by the triangles. The theory results from the three-level model are indicated via the solid lines, of which the opacity is set by the overlap of the states with the initial polaron state.   
    The horizontal black markers in (a) denote the lowest energies for which nonzero rf transfer was identified.}
\end{figure}

In the experimental spectrum, we do not observe the isolated lines from the three-level model, but a broad continuous feature.
That is to be expected, since both the dimer and trimer can scatter off particles of the BEC or deform the BEC, leading to a continuum of gapless excitations which is probed in our spectroscopy method. Line-of-sight integration and three-body loss contribute to additional broadening of the resonances. Moreover, our approach still underestimates the lowest attractive polaron energy. This is likely due to the admixture of states with an even higher number of excitations from the medium. For instance, the dimer-trimer superposition is likely to be itself dressed by its own attractive polaron cloud on top of the included effects. This dressing may also include coupling to a possible bound tetramer state. While we have not observed evidence for such a tetramer in the thermal gas~\cite{chuang:2025}, we cannot exclude its existence.

We now study the polaron rf-spectrum as a function of the interaction strength between the impurity in the final state $\ket{\uparrow}$ and the bath, shown in Fig.~\ref{fig:E_vs_B}.
The theory curves (opaque lines) are computed using the three-level model discussed before, where we account for the initial-state energy.
We find that the two lowest eigenenergies of the theoretical three-level model reliably describe the width of the attractive polaron branch over the wide range of interaction strengths explored. Indeed, in the entire range of magnetic fields probed, the energy splitting between the dimer and trimer remains small ($<$25 kHz)~\cite{chuang:2025}, and dimer-trimer hybridization persists for all interaction strengths. A full theory of spectral line shapes in the presence of strong interactions and three-body correlations, including the repulsive polaron, is an open problem beyond the scope of this work.

We note that for fields above the final state impurity-boson Feshbach resonance at $B{\approx}110\,\rm G$~\cite{park:2012}, the scattering lengths of initial and final spin states approach each other and are large or comparable to the interboson distance ${\sim}1/k_n$. Despite strong initial- and final-state interactions, the rf spectrum then becomes narrow~\cite{Gupta2003,Baym2007,Zwierlein2016a}.

\begin{figure}[t]
	\includegraphics[width=0.95\columnwidth]{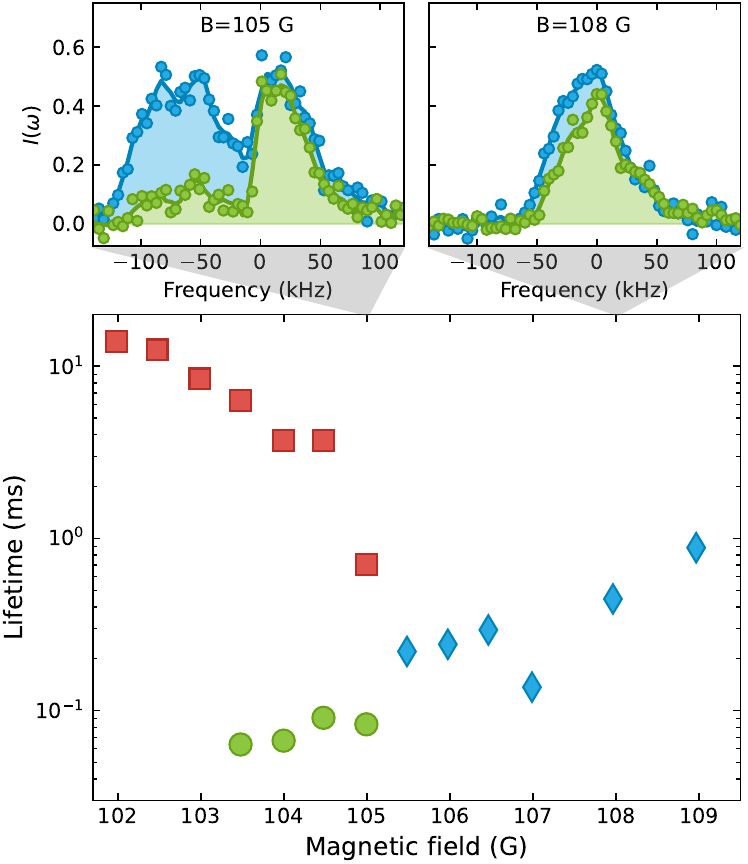}
	\caption{\label{fig:lifetime} Lifetimes of the states created by rf injection. These lifetimes are inferred from the ratio of population depleted from $\ket{\downarrow}$ to population observed in $\ket{\uparrow}$ (upper panels and Fig.~\ref{fig:E_vs_B}). For data taken at $B\leq105\,\rm G$, we resolve the attractive and repulsive branches of the spectrum, whereas for larger magnetic fields the two features merge (see upper panels). The lifetime of the attractive polaron states (green circles) decreases at deeper binding energy,
    whereas the lifetime of the repulsive polaron branch (red squares) increases. Blue diamonds: lifetime of the merged feature above $105\,\rm G$. At sufficiently deep binding energies, we do not observe any population in $\ket{\downarrow}$.}
\end{figure}

The difference between the depletion and arrival spectra in Fig.~\ref{fig:E_vs_B} can be attributed to the finite lifetime of the states created by the rf injection. Three-body recombination leads to a release of kinetic energy and the ejection of the involved atoms from the trap~\cite{wolf:2017}. By comparing the number of atoms observed in the final state to the number of atoms depleted in the initial state, we can estimate the decay lifetimes $1/\Gamma$ of the states prepared during the rf injection pulse time. We model the transfer between populations $P_\downarrow, P_\uparrow$ as $dP_{\downarrow} \propto -\Omega_R(t)^2 P_{\downarrow} dt$, where the Rabi frequency $\Omega_R$ is time dependent due to pulse shaping. The time-dependence of the $\ket{\uparrow}$-population is modeled as $dP_{\uparrow} = -dP_\downarrow - \Gamma P_\uparrow dt$. The lifetimes thus obtained are shown in  Fig.~\ref{fig:lifetime}.

While in the repulsive polaron branch the depletion and arrivals signals are comparable, in the attractive branch the arrival signal is much weaker. This points to a shorter lifetime of the attractive polaron, which is consistent with a large trimer admixture to the wave function, since trimers have a limited lifetime due to internal three-body recombination. Moreover, we find that the lifetime decreases substantially with increasing binding energy. Qualitatively, this is to be expected, since the probability that the three particles meet each other at short distance and undergo the recombination reaction increases as the trimer shrinks for larger binding energy.

At the magnetic fields where we are able to detect nonzero population in $\ket{\downarrow}$ in the attractive branch, we observe decay lifetimes $1/\Gamma$ greater than $50\,\mu s$. Up to factors of order unity, this contributes to a lifetime broadening of $\Gamma/(2\pi) \approx 3$ kHz, whereas the typical spectral widths measured ($>$ 50 kHz, e.g., Fig. \ref{fig:lifetime} top left) are at least an order of magnitude larger. This confirms that the broad spectra we observe cannot simply be attributed to loss and further supports the interpretation based on BEC-induced hybridization of the dimer and trimer states.


In conclusion, we have observed conceptually novel molecular states: superpositions of dimers and trimers arising naturally by polaronic coupling to a quantum degenerate bosonic bath. We find that the dimer-trimer mixing induced by the BEC causes strong level repulsion, setting one of the dominant energy scales in our spectra. We show that by preparing an initial attractive polaron state, we are able to explore the crossover between few- and many-body physics of fermions immersed in a BEC within experimentally meaningful timescales. Furthermore, our measurements provide an excellent testbed to further develop the challenging theoretical descriptions of Bose polarons~\cite{grusdt:2025,massignan:2025} and more generally Bose-Fermi mixtures~\cite{viverit:2002,kinnunen:2018,vonmilczewski:2022,duda:2023,shen:2024}. 

Beyond their significance for polaron physics, our results  connect to chemistry: recent works have shown that coherence in a BEC can strongly affect reaction dynamics~\cite{zhang:2023,nagata:2025}. Our work shows that in ultracold mixtures even richer effects can be expected, where the fundamental identity of molecules is reshaped. This opens up an exciting new frontier on the intersection between quantum many-body physics and ultracold chemistry~\cite{krems:2008,karman:2024}. In this field, control of chemical reactions on the two-body level has been achieved~\cite{ospelkaus:2010,liu:2022,liu:2024}, but the effect of a surrounding medium has not been widely discussed, although in conventional chemistry solvents play a critical role ~\cite{reichardt:2010}.

\textbf{Acknowledgements.---}We thank  Meera Parish and Jesper Levinsen for inspiring discussions, and Yiqi Ni for early contributions to the experiment.
This work was supported by the NSF through the Center for Ultracold Atoms,
PHY-2012110 and PHY-2513210, AFOSR through FA9550-23-1-0402 and a MURI on Ultracold Molecules, and the Vannevar Bush Faculty Fellowship (ONR N00014-19-1-2631). A.~C. acknowledges funding from an ETH fellowship.
R.~S. acknowledges support by the Deutsche Forschungsgemeinschaft under Germany's Excellence Strategy EXC 2181/1 - 390900948 (the Heidelberg STRUCTURES Excellence Cluster), the CRC 1225 ISOQUANT, project-ID 273811115, and the DFG scientific network `A(E)MP - Appearance of the Effective Mass in Polaron Models' (Grant No. 569490025).

\section{Supplementary Material}

\subsection{Spectroscopy and Detection}

\begin{figure}[t]
    \centering
    \includegraphics[width=0.98\columnwidth]{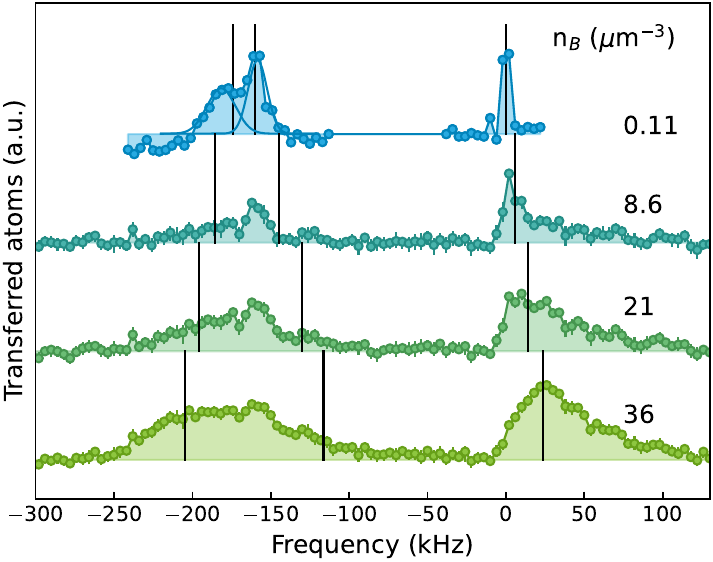}
    \caption{The population transfer from impurity spin state $\ket{\downarrow}$ at $B=102.5\,\rm G$ measuring the remaining population in $\ket{\downarrow}$ as a function of frequency for several densities of the bosonic bath. Note the double peak structure corresponding to the dimer and trimer feature for the lowest density spectrum, taken with a thermal, uncondensed bosonic cloud. The solid lines correspond to the theory results from the three-level model.}
    \label{fig:expcuts}
\end{figure}

 To perform rf injection, we drive with frequencies $\sim 26\,\rm MHz$ using a broadband, in-vacuum antenna, providing a Rabi frequency of $\sim 2\pi \times 10\,\rm kHz$ between the two lowest hyperfine states of $^{40}$K, denoted $\ket{\downarrow}$ and $\ket{\uparrow}$. To suppress the spectral sidelobes beyond the Fourier window $1/T_{\rm rf} < 4\,\rm kHz$ due to a finite rf pulse time $T_{\rm rf}\geq 275\,\rm \mu s$, we amplitude modulate the pulse with a Blackman window function, with parameter $\alpha=0.16$. The Rabi frequency in the absence of the condensate as a function of time is then $\Omega_{\rm R}(t) = \Omega_{\rm max} ((1-\alpha)/2 - \cos(2 \pi t/T_{\rm rf})/2 + \alpha \cos(4 \pi t/T_{\rm rf})/2)$. The pulse time, ranging from 275 to 1600 $\mu$s, is varied  with the impurity-bath interaction strength, such that $\sim$50\% of the initial polaron $\ket{\downarrow}$ population is injected into the manifold of bound states. Adapting the pulse time is necessary to obtain a clear attractive polaron signal, as the coupling strength between the initial attractive polaron and bound states changes strongly near the Feshbach resonance. Subsequently, populations are measured via absorption imaging. Example line cuts of the experimental spectra in Fig.~\ref{fig:halfMoon} are shown in Fig.~\ref{fig:expcuts}.

 To measure the populations in each spin state $\ket{\downarrow}$ and $\ket{\uparrow}$ we perform on-resonant absorption imaging, using the corresponding $\sigma_+$ transitions to the $^{40}$K $P_{3/2}$, $F'=11/2$ excited state hyperfine manifold. The probe light for each of the two states is also sensitive to the population in the other state, due to off-resonant scattering. For our relevant experimental parameter regime, the detuning to linewidth ratio is $\Delta/\Gamma \approx 17\,\mathrm{MHz}/6\,\mathrm{MHz} \approx 2.8$, which corresponds to an off-resonant scattering cross-section of $\sigma = \sigma_0 / (1 + 4(\Delta/\Gamma) ^ 2) \approx 0.03 \,\sigma_0$, for on-resonant cross-section $\sigma_0$.  We avoid contaminating the measurement of small populations in $\ket{\uparrow}$ (fraction $\lessapprox$ 0.1) by shelving $\ket{\uparrow}$ to an auxiliary ground hyperfine state $\ket{F=7/2, m_F=-7/2}\equiv \ket{\mathrm{aux}}$ prior to imaging the population in $\ket{\uparrow}$.
 Afterwards, the population in $\ket{\mathrm{aux}}$ is returned to $\ket{\downarrow}$ for imaging. This shelving procedure is designed to cleanly probe small (fraction $\lessapprox$ 0.1) populations in $\ket{\uparrow}$, whereas measuring the typical populations in $\ket{\downarrow}$ (fraction $\gtrapprox$ 0.5) is much less demanding.

The $1/e$ lifetime of the attractive polaron state is only tens of $\mu$s, constraining the time available for the shelving procedure. Thus, we implement shelving within 10$\mu$s by using an optical Raman transition via the excited $^{40}$K $P_{1/2}$ state, with a maximum two-photon Rabi frequency $\gg 500\,\rm kHz$, and one-photon detuning $\approx$ 20 GHz. Both the pump and Stokes coupling derive from a single spatial mode near-resonant beam propagating normal to the quantization B-field, with their two frequencies satisfying the two-photon resonance condition between $\ket{\uparrow}$ and $\ket{\mathrm{aux}}$. 

The optical shelving pulse is configured for insensitivity to pulse timing and two-photon resonance frequency shifts. We adiabatically chirp the two-photon detuning over 2.5 MHz and amplitude modulate the two-photon Rabi frequency with a Blackman window function. This procedure maps to a rapid adiabatic passage in the approximate two-level system spanning $\ket{\uparrow}, \ket{\mathrm{aux}}$ \cite{vitanov:2017}. The round-trip fidelity of shelving and returning the population was measured to be 0.99 by probing the population in $\ket{\downarrow}$ after up to five round-trips, in the absence of the bath.

\subsection{Theoretical modeling}

We consider the problem of a mobile impurity of mass $M$ in a homogeneous BEC of bosons with mass $m$ and chemical potential $\mu_{\rm B}$. We denote the impurity-boson interaction potential by $V_{\rm IB}$ and the boson-boson interaction potential by $V_{\rm BB}$. The interactions are treated within a single-channel model, which is justified since the boson-impurity scattering length is tuned close to a broad Feshbach resonance. We treat the impurity in first quantization with quadrature operators $\hat{\bm{R}}$ and  $\hat{\bm{P}}$, and the bosons in second quantization with creation and annihilation operators $\hat{b}^{\dagger}_{\bm{k}}$ and $\hat{b}_{\bm{k}}$, respectively. We set $\hbar=1$. This yields the Hamiltonian
\begin{multline} \label{eq:origHam}
    \hat{\mathcal{H}}_0=-E_{\rm bg}+\int_{\bm{k}} \Bigl(\frac{k^2}{2m}-\mu_{\rm B} \Bigr) \hat{b}^{\dagger}_{\bm{k}} \hat{b}_{\bm{k}}+\frac{\hat{\bm{P}}^2}{2M} \\ +\int_{\bm{r}} V_{\rm IB}(\bm{r}-\bm{\hat{R}}) \hat{b}^{\dagger}_{\bm{r}} \hat{b}_{\bm{r}}+ \frac{1}{2} \int_{\bm{r'}} \int_{\bm{r}} V_{\rm BB}(\bm{r'}-\bm{r}) \hat{b}^{\dagger}_{\bm{r'}} \hat{b}^{\dagger}_{\bm{r}} \hat{b}_{\bm{r'}}  \hat{b}_{\bm{r}}.
\end{multline}

Here $E_{\rm bg}$ is the energy of the background BEC without the impurity. We denote $\int_{\bm{r}}=\int d^3r$ and $\int_{\bm{k}}=\frac{1}{(2\pi)^3}\int d^3k$. The chemical potential is set to the mean field value $\mu_{\rm B}=n_0 U_{\rm B}$, where $n_0$ is the density of the BEC and $U_{\rm B}$ the coupling constant on the level of the Born approximation.

\begin{figure}
\includegraphics[width=\linewidth]{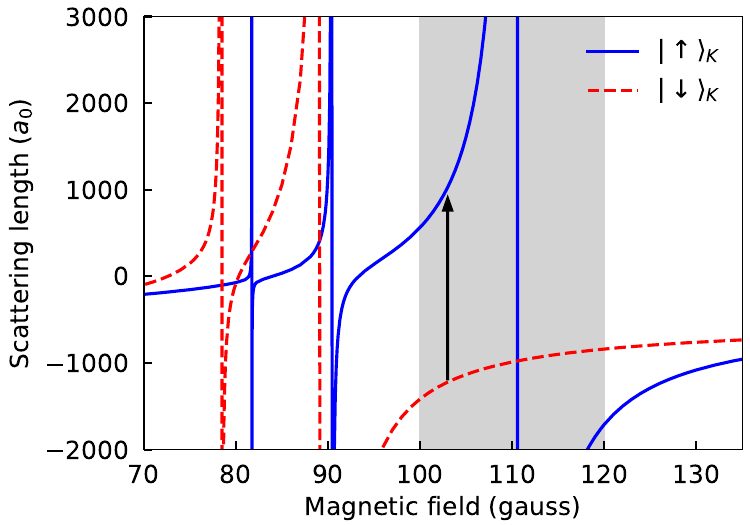}
\caption{Adapted from Ref.~\cite{chuang:2025}. Scattering length (in $a_0$)  of scattering between $^{23}$Na
and $^{40}$K as a function of magnetic field, for the $\left|\downarrow\right\rangle$ (red dashed) and $\left|\uparrow\right\rangle$ (blue solid) states of K. The gray shaded area indicates the experimentally probed magnetic field range.}
\label{fig:Feshbachplots}
\end{figure}
\subsubsection{Potential calibration}
Our modeling for the NaK and Na-Na potentials ($V_{\rm IB}$ and $V_{\rm BB}$) is described in Ref.~\cite{chuang:2025}. In brief, we compute the magnetic-field-dependent scattering length and effective range with coupled channels calculations using the realistic potentials from Ref.~\cite{hartmann:2019}. The scattering lengths of the initial and final states are shown in Fig.~\ref{fig:Feshbachplots}. In the experimentally probed regime of magnetic fields, the dimer energy is well described using the universal expression
\begin{equation}
E_{\mathrm{dim}}=-\frac{1}{2 \mu r_{\mathrm{eff}}^2} \left(1-\sqrt{1-\frac{2 r_{\mathrm{eff}}}{a}}\right)^2,
\end{equation}
where the effective range contribution is important to obtain quantitative agreement \cite{chuang:2025}. In the experimental magnetic field regime, the initial state attractive polaron energy at the maximal experimental density ranges from -11 to -7 kHz.

To make our calculations tractable, we replace the physical interaction potential between the Na and K-atoms by a single-channel Gaussian model potential 
\begin{equation}
V_{\rm IB}(\bm{r})=\frac{g}{2L_g^2}\exp \left(-\frac{r^2}{L_g^2} \right),
\end{equation}
with the same scattering length and effective range \cite{christianen:2024}. Here $g$ sets the coupling strength and $L_g$ the range of the potential. We do this separately for every magnetic field. We use the same procedure to construct the Gaussian model potentials for the initial state of the spectroscopic protocol.

Similarly, we compute the scattering length and effective range for the Na-Na scattering, using potentials from~\cite{knoop:2011}. Over the magnetic field range of interest we find that the scattering length is approximately $56.5\, a_0$ with variation of only $0.2 a_0$, which we neglect. In this case there does not exist a single Gaussian potential without bound states with the correct combination of scattering length and effective range. Therefore, we use a sum of two Gaussian potentials with coupling strengths and ranges $U_i$ and $L_{U,i}$ respectively. The precise procedure for constructing the potentials and the final parameters are given in Refs.~\cite{christianen_thesis:2023,chuang:2025}.

With these potentials we obtain good agreement for the trimer energy with the experimental results \cite{chuang:2025}. There is a systematic discrepancy in the dimer-trimer gap of approximately 5 kHz, but this is comparable with the size of the experimental error bar for systematic shifts.

\subsubsection{Variational procedure}

We use the variational framework from Ref.~\cite{christianen:2024} to obtain the ground-state energy. We treat the background BEC on the mean-field level and displace the Hamiltonian using the unitary displacement operator. We then apply the Lee-Low-Pines transformation to go to the reference frame of the impurity. In this reference frame, we consider the double-excitation Ansatz for the wavefunction of the bosons
\begin{multline}
|\psi[\beta_0,\beta(\bm{r}),\alpha(\bm{r},\bm{r}')]\rangle=\beta_0+ \int d^3r \beta(\bm{r}) \hat{b}^{\dagger}_{\bm{r}} \\+ \frac{1}{\sqrt{2}} \int \int d^3r d^3r' \alpha(\bm{r},\bm{r}')  \hat{b}^{\dagger}_{\bm{r}}  \hat{b}^{\dagger}_{\bm{r}'} |0\rangle.
\end{multline}
Here $\hat{b}^{\dagger}_{\bm{r}}$ is the bosonic creation operator, creating a boson at position $\bm{r}$. The vacuum state $|0\rangle$ here corresponds to the state of the BEC with the impurity at the center of the reference frame. We express the variational parameters $\beta(\bm{r})$ and $\alpha(\bm{r},\bm{r}')$ as sums of (products of) spherical Gaussian basis functions, taking into account the spherical symmetry of the problem to reduce the computational complexity.

\subsubsection{Treatment of the interboson repulsion}

The most challenging part of the variational calculation is the treatment of the interboson repulsion. This is discussed in detail in Ref.~\cite{christianen:2024}. The Hamiltonian term corresponding to the interboson repulsion, after displacing the background BEC, is given by
\begin{multline} \label{eq:Ham_rep}
\hat{\mathcal{H}}_U=\sum_{i=1,2} \int \int d^3r' d^3r \frac{U_i}{2L_{Ui}^2}\exp[-\frac{(\bm{r'}-\bm{r})^2}{L_{Ui}^2}] \\  [\frac{n_0}{2}  (2\hat{b}^{\dagger}_{\bm{r'}} \hat{b}_{\bm{r}} +\hat{b}^{\dagger}_{\bm{r'}} \hat{b}^{\dagger}_{\bm{r}}+ \hat{b}_{\bm{r'}} \hat{b}_{\bm{r}}) \\+ \sqrt{n_0} (\hat{b}^{\dagger}_{\bm{r'}} \hat{b}^{\dagger}_{\bm{r}}  \hat{b}_{\bm{r}}+ \hat{b}^{\dagger}_{\bm{r}} \hat{b}_{\bm{r'}}  \hat{b}_{\bm{r}})+ \frac{1}{2}  \hat{b}^{\dagger}_{\bm{r'}} \hat{b}^{\dagger}_{\bm{r}} \hat{b}_{\bm{r'}}  \hat{b}_{\bm{r}}].
\end{multline}
To obtain the correct trimer energy, the interboson repulsion needs to be taken fully into account when describing the trimer wave function, \textit{i.e.}, for the last term in Eq.~\eqref{eq:Ham_rep}. However, accurately describing the interboson repulsion in the entire BEC is difficult in our variational wave function. Therefore, we treat the interboson repulsion in the BEC on the level of the Born approximation. In practice, we replace the coupling strengths $U_i$ by a rescaled coupling strength $U^{(B)}_i$ for any term in the Hamiltonian where the background BEC has been displaced. The value of $U^{(B)}_i$ is set so that it gives the correct scattering length on the level of the Born approximation. The modified version of the interboson repulsion Hamiltonian is thus given by 
\begin{multline}
\hat{\mathcal{H}}_U=\sum_{i=1,2}\frac{1}{2L_{Ui}^2} \int \int d^3r' d^3r \ U^{(B)}_i\exp \left[ -\frac{(\bm{r'}-\bm{r})^2}{L_{Ui}^2} \right] \\  \left[2 n_0\hat{b}^{\dagger}_{\bm{r'}} \hat{b}_{\bm{r}}+ \sqrt{n_0} (\hat{b}^{\dagger}_{\bm{r'}} \hat{b}^{\dagger}_{\bm{r}}  \hat{b}_{\bm{r}}+ \hat{b}^{\dagger}_{\bm{r}} \hat{b}_{\bm{r'}}  \hat{b}_{\bm{r}})+ \frac{1}{2}  \hat{b}^{\dagger}_{\bm{r'}} \hat{b}^{\dagger}_{\bm{r}} \hat{b}_{\bm{r'}}  \hat{b}_{\bm{r}}\right] \\
+ \frac{(U_i-U^{(B)}_i)}{2} \exp \left[-\frac{(\bm{r'}-\bm{r})^2}{L_{Ui}^2}-\frac{\bm{r'}^2+\bm{r}^2}{L_{W}^2} \right] \hat{b}^{\dagger}_{\bm{r'}} \hat{b}^{\dagger}_{\bm{r}} \hat{b}_{\bm{r'}}  \hat{b}_{\bm{r}}.
\end{multline}
Aside from the changes in the interaction constants, in this expression we have replaced the quadratic terms $\hat{b}^{\dagger} \hat{b}^{\dagger}$ and $\hat{b} \hat{b}$ by $\hat{b}^{\dagger} \hat{b}$ to avoid renormalization of the scattering in the background BEC. Indeed, since we now treat the interactions on the bath on the level of the Born approximation, the variational Ansatz should not be allowed to treat these interactions on a higher level. With this form of the Hamiltonian the correct mean-field value of the polaron energy is recovered. 

Finally, with the expansion in spherical Gaussian basis functions it is difficult to converge the interboson interactions when the bosons are both far away from the impurity. To resolve this, we model the interboson repulsion as a three-body interaction which reduces to the unmodified potential close to the impurity, and the Born-approximation result far from the impurity. The length-scale $L_W$ separating these regimes, can be calibrated by comparing the result to an accurate computation of the trimer energy. For more details, see Refs.~\cite{christianen_thesis:2023,chuang:2025}.

Altogether, the general philosophy of this approach is to describe accurately the interboson interactions close to the impurity, while addressing computational difficulties by treating the interboson interactions on the level of the Born approximation far from the impurity.

\subsubsection{Three-level model}

For the interpretation of the spectra, we use the variationally obtained wave function as an input for the three-level model.
 The three-basis functions of the three-level model are taken to be
\begin{align}
|1 \rangle &= |0 \rangle, \\
|2 \rangle &= \int d^3r \beta(\bm{r}) \hat{b}^{\dagger}_{\bm{r}} |0 \rangle, \\
|3 \rangle &=\frac{1}{\sqrt{2}} \int \int d^3r d^3r' \alpha(\bm{r},\bm{r}')  \hat{b}^{\dagger}_{\bm{r}}  \hat{b}^{\dagger}_{\bm{r}'} |0\rangle.
\end{align}
The three-level Hamiltonian is found by computing the matrix elements of the Hamiltonian with respect to these three states. By construction, the ground-state energy in this three-level model is the ground-state energy also found from the variational minimization of the double-excitation Ansatz. The coefficients on the off-diagonal of the matrix in Fig.~\ref{fig:intro}(c) are given by
\begin{align}
    g_1&=\int d^3 r V(r) \ \beta(r), \\
    g_2&=\sqrt{2} \int \int d^3 r d^3r' \ \alpha(r,r') \beta(r) V(r').
\end{align}

\begin{figure}[t!]
\includegraphics[width=0.8\columnwidth]{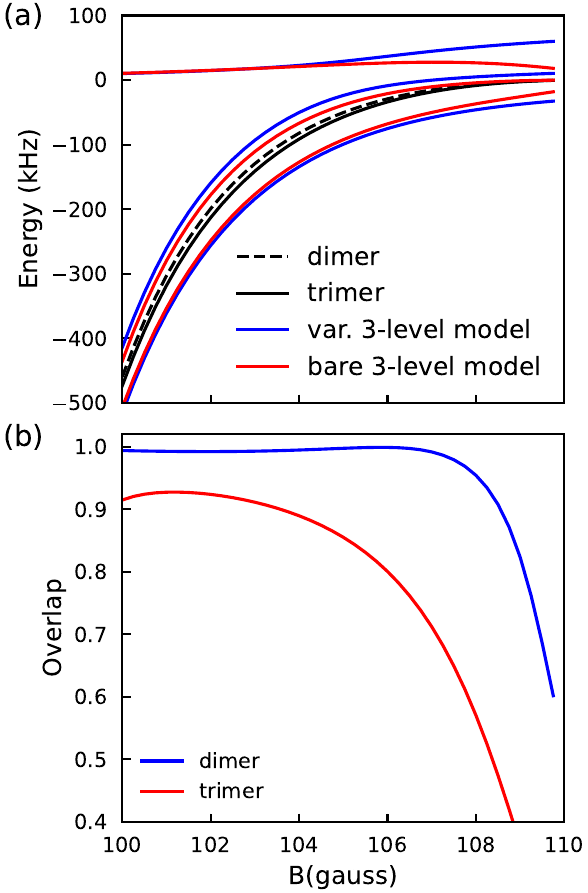}
\caption{Comparison of the three-level model with the bare dimer and trimer wave functions, and the model with the dimer and trimer wave functions obtained from the variational double-excitation approach. (a) shows the eigenvalues of the three-level model as a function of magnetic field, and (b) shows the overlap between the bare and the variational dimer and trimer wave functions.}
\label{fig:bare_vs_dressed}
\end{figure}

Since the wave functions $\beta$ and $\alpha$ are optimized in  presence of the BEC, it is not a priori clear that they resemble the original dimer and trimer wave functions. To show that the picture of dimer-trimer superpositions is well justified, we compare the variational three-level model just described with the bare three-level model, where the wave functions for level $|2\rangle$ and $|3\rangle$ are replaced by the dimer and trimer wave functions in the vacuum. In Fig.~\ref{fig:bare_vs_dressed}(a) we compare the eigenenergies found from the three-level models, and in Fig.~\ref{fig:bare_vs_dressed}(b) we compare the bare and variational dimer and trimer wave functions. We consider the highest experimental BEC densities of 4$\times 10^{13}$cm$^{-3}$.

In Fig.~\ref{fig:bare_vs_dressed}(a) we find that the energies of the three-level model are quantitatively different in the full microscopic calculation, but that even the three-level model with the bare dimer and trimer states captures the key physics. In presence of the BEC the dimer and trimer wave functions are modified, so that the coupling with the BEC is stronger. As a result, the variational solution has a lower ground-state energy and a larger level repulsion. 

In Fig.~\ref{fig:bare_vs_dressed}(b) we show the overlaps between the bare $|i\rangle_b$ and variational basis states $|i\rangle_v$, given by $|_b\langle 2|2\rangle_v|^2$ and $|_b\langle 3|3\rangle_v|^2$. The overlaps are large especially away from the resonance. When the resonance is approached, the dimer-trimer splitting becomes much smaller than the coupling with the BEC and the description in terms of the bare dimer and trimer states breaks down. This is expected, since the ground-state energy connects continuously with the attractive polaron state at negative scattering lengths, where the bare dimer and trimer states do not exist. Our analysis thus supports the interpretation of the experimentally observed signal as originating from the hybridization between the dimer and trimer states, especially far from the resonance.

For the direct comparison with the experimentally observed spectrum, the initial-state polaron energy needs to be subtracted from the energies obtained from the variational three-level model. This procedure results in the theoretical lines in Figs.~\ref{fig:halfMoon} and \ref{fig:E_vs_B}. In Fig.~\ref{fig:E_vs_B} the opacity of the lines reflects the overlaps of the states in the three-level model with the initial state. Since the two-and three-body components of the initial-state wave function are different from the final-state wave function, these weights do not sum up to 100\%. 

Nevertheless, we still find that the weights add up to 80\% - 99\%, depending on the magnetic field and the density. This means that the three levels together can capture well the wave function of the initial state.

In principle one can extract a whole spectrum for the Hilbert space truncated to two excitations. However, since there are general limitations to this approach we do not believe it to be more insightful than the three-level model. Indeed, the double-excitation Ansatz misses effects which are not crucial for the ground-state energy but more important for the spectrum. These include dressing of the molecular state by a long-range polaron cloud or phononic excitations in both initial and final state. Furthermore, no three-body loss is included.

The calculation of the full spectrum which can be matched to the experiment thus remains an interesting open challenge. However, since the ground-state energy, the width of the spectrum, and the distribution of spectral weight are already well reproduced using the three-level model, we believe that the three-level mixing is the most essential ingredient in the description.

\end{document}